\documentclass[aps,prl,twocolumn,superscriptaddress,amsmath,amsfonts,showpacs]{revtex4}

\usepackage{graphicx}

\begin{document}


\title{Zero-bias conductance peak split in $d$-wave superconductors: \\
Derivation of the universal magnetic field dependence}


\author{Christian Iniotakis}
\affiliation{Institute for Theoretical Physics,  ETH Zurich, 8093 Zurich, Switzerland}


\email[]{iniotakis@itp.phys.ethz.ch}


\date{\today}

\begin{abstract}
The zero-bias conductance peak in $d$-wave superconductors splits in an applied magnetic field. 
In this work, the experimentally observed universal relation $\delta \propto \sqrt{B_0}$ for strip-shaped
samples is derived analytically based on the long-ranged current contributions from Abrikosov vortices
inside the sample. The result is in full agreement with observed key properties, and features such as hysteresis
effects are made accessible. Employing a magnetically induced additional order parameter is not necessary
for the physical explanation of the universal relation.
\end{abstract}



\maketitle


It is well known, that the surface of a material can display a different physical behavior than the bulk. 
Nevertheless,  a thorough analysis of these surface effects sometimes
reveals important and unique information about the bulk system itself. 
An amazing example are fermionic states,
commonly referred to as Andreev bound states, at the surface of unconventional
superconductors \cite{Hu,Tanaka95,Buchholtz,Covington,KashiwayaReport,Iguchi,Yamashiro,Honerkamp,Laube,Mao,IniotakisNCS}. 
Living on a surface sheet of a few coherence lengths  only,
these Andreev bound states are responsible for significant zero-bias anomalies in tunneling experiments: 
They appear in both singlet high-$T_c$ cuprates \cite{Hu,Tanaka95,Buchholtz,Covington,KashiwayaReport,Iguchi} and triplet superconductors 
such as Sr$_2$RuO$_4$ \cite{Yamashiro,Honerkamp,Laube,Mao} and they are also expected to form in non-centrosymmetric superconductors 
of mixed parity like CePt$_3$Si \cite{IniotakisNCS}.  For the fundamental question of identifying the superconducting gap function 
of a given bulk material, the surface bound states have proven to be of great importance \cite{Hu,Tanaka95,Buchholtz,Covington,KashiwayaReport}. 
Their mere existence is a clear indication, that  the pairing interaction underlying the superconductivity in the sample is not of the conventional
isotropic $s$-wave type. Moreover, detailed measurements of the zero-bias anomalies for different surface 
orientations provide helpful information about the actual gap structure in momentum space \cite{KashiwayaReport,Iguchi}.
Apart from their existence and orientational dependence, another valuable source of information is based on the behavior of the zero-bias anomalies 
under the influence of an applied magnetic field. As an example, the chirality of a $p$-wave superconducting gap function 
might be detected due to the unique response of the surface bound states, 
 reacting differently on magnetic fields of the same magnitude, but antiparallel orientation  \cite{TanakaMagnet,Yokoyama}. 
For a $d$-wave superconductor such as YBa$_2$Cu$_3$O$_{7-x}$, which is in the focus
of this work, the conductance peak at zero bias is known to split into two peaks at finite energies $\pm \delta$ when a 
magnetic field is applied parallel to the $c$ axis \cite{CovingtonPRL,Aprili,Krupke,Dagan}.  
Whereas the experimental evidence for the splitting is clear and fully established already for quite a long time, the question
about its physical origin is still open. 
The main reason for this is the existence of different physical mechanisms, which are capable of explaining this behavior in general.
First of all, a natural explanation for the splitting is a Doppler shift of the Andreev bound state energies due to the superfluid velocity
of the condensate at the surface \cite{Fogelstroem}. 
 It is important to realize, however, that due to the Doppler shift  not only Meissner 
currents screening the applied magnetic field contribute to the total splitting,  but also Abrikosov vortices from 
inside the superconductor, since the currents surrounding them are long-ranged  \cite{Graser, Iniotakis}. 
 A different mechanism, which would also lead to a splitting of the zero-bias conductance peak, is the existence of a small 
minority order parameter component of the $i d_{xy}$ or $i s$ type in addition to the main $d_{x^2-y^2}$ order parameter.
For this scenario, the magnitude of the observed split should be directly proportional to the magnitude of the 
minority order parameter component \cite{Tanuma}.

A very important quantitative observation, which can be of great help in identifying the origin of the zero-bias 
conductance peak splitting, has been established in a series of experiments done by the Deutscher group 
\cite{Beck, Elhalel, Leibovitch}. 
Performing tunneling measurements on various narrow, strip-shaped samples of YBa$_2$Cu$_3$O$_{7-x}$, they
found out, that the zero-bias conductance peak split for decreasing magnetic fields obeys the law 
\begin{equation}
\label{EQSqrtB}
\delta=\alpha \sqrt{B_0}.
\end{equation}
Here, $B_0$ denotes the magnitude of the applied magnetic field, and  $\alpha$ can be determined from experiment. 
Interestingly, the asymptotic behavior of all the samples is universally described by Eq. (\ref{EQSqrtB}) with 
$\alpha \approx 1.1 \textrm{meVT}^{\frac{1}{2}}$ \cite{Beck,Leibovitch}. 
There is a low-field deviation, which seems to correlate to the individual doping level of the sample, but already
for intermediate magnetic fields the above relation universally holds true, independent of the specific doping.
The width of the measured samples varies between 600 {\AA} and 3200 {\AA}. Amazingly, it does not have 
any influence, either.
The only exception reported sofar is a sample, which is considerably smaller in width, and fulfils Eq. (\ref{EQSqrtB})
with $\alpha \approx 0.65 \textrm{meVT}^{\frac{1}{2}}$ \cite{Elhalel}. 
Up to now, the experimental observation of the universal relation has often been linked to a
theoretical model proposed by Laughlin \cite{Laughlin}. This model assumes
the formation of an additional $i d_{xy}$ order parameter in the bulk of the 
superconductor, which should be induced by an applied magnetic field.
In this scenario, the magnitude of the induced order parameter, and thus
also the corresponding energy split, would obey a relation of the same 
non-linear type as Eq. (\ref{EQSqrtB}), namely $\delta \propto \sqrt{B_0}$. 
It is important to note, however, that Laughlin's model does not at all include
the contribution of Abrikosov vortices in the sample.
Since the universal relation Eq. (\ref{EQSqrtB}) 
 is measured for applied 
magnetic fields of up to 16 T, which are generally much higher than the lower critical field,
 there are undoubtedly huge numbers of Abrikosov vortices  inside the samples.
Therefore, it is essential to properly take into account their contribution to the superfluid
velocity at the surface. 
In the following it will be clarified, that the presence of Abrikosov vortices
already results in the universal relation Eq. (\ref{EQSqrtB}) and explains its key features.
In particular, it is not necessary to rely on an additional magnetically induced order parameter.
  
\begin{figure}[t]
\includegraphics[width=0.65 \columnwidth]{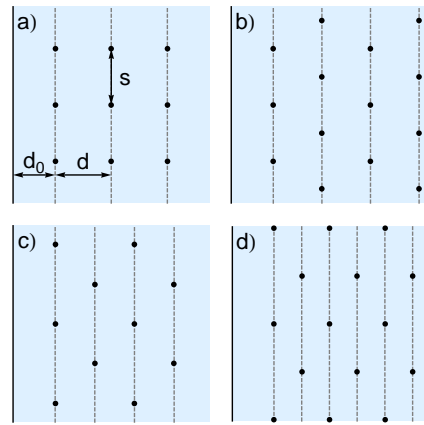}
\caption{\label{Fig01} (color online)
Sketch of different regular vortex lattices situated at an offset  $d_0$ from the
boundary of the superconductor. Each of the lattices can be characterized by 
specific lines, which are parallel to the boundary (dashed). Along such a line, adjacent vortices have
 a distance $s$, and two neighbouring lines themselves are separated by the distance $d$. 
In units of the intervortex distance between nearest neighbours, we have a) quadratic lattice: $s=d=1$, 
b) triangular lattice: $s=1$, $d=\sqrt 3/2$, c) rotated quadratic lattice: $s=\sqrt 2$, $d=1/\sqrt 2$,
d) rotated triangular lattice: $s=\sqrt 3$, $d=1/2$. The ratio $q_\Lambda=d/s$ depends on both 
lattice type and orientation.
}
\end{figure}

As a starting point for the derivation, we consider a $d$-wave superconducting sample in an
applied magnetic field $B_0 \gg B_{c1}$, so that Abrikosov vortices are expected inside naturally. 
It is beyond the scope
of this work to calculate the positions of all these vortices ab initio. Rather, it is assumed
for simplicity, that the Abrikosov vortices form a regular lattice $\Lambda_0$, which
is situated at an offset  $d_0$ from the boundary. A stability condition ensuring
the physical relevance of this model is employed at a later stage.
Characteristic samples of such an Abrikosov lattice are depicted
in Fig.~\ref{Fig01}, a)-d), for two different lattice types, namely 
quadratic and triangular ones, at different orientations. The lattices in Fig. \ref{Fig01}, a) and b), 
have a connection between
nearest neighbours, which is parallel to the boundary, whereas those in c) and d) are rotated.
It is important to note, that we may find lines parallel to
the boundary (marked as dashed), which intersect all vortices in such a way, 
that two adjacent vortices on such a line always have the same
distance $s$. Moreover, the spacing between these lines 
is equidistant and  denoted by $d$ in the following. Obviously, the 
area of a unit cell of the lattice is given by the product $s\cdot d$, whereas
the ratio $q_\Lambda=d/s$ depends on both the lattice type and its orientation, which will
have consequences on the resulting zero-bias conductance peak 
split as shown below.

\begin{figure}[t]
\includegraphics[width=0.75 \columnwidth]{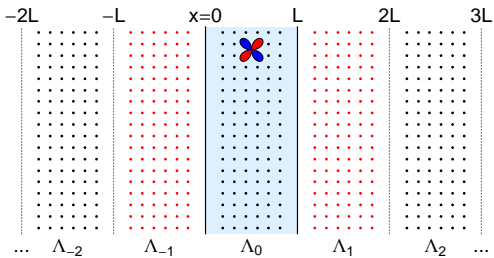}
\caption{\label{Fig02} (color online)
Sketch of a regular vortex lattice $\Lambda_0$ inside the $d$-wave superconducting strip (blue)
of width $L$. The orientation of the $d$ wave is indicated by the lobes. 
To ensure the correct implementation of electrodynamic boundary conditions, an infinite 
number of virtual mirror lattices $\Lambda_{n\neq0}$ has to be taken into account. 
For odd $n$, $\Lambda_n$ consists of antivortices (red). 
}
\end{figure}

Abrikosov vortices are surrounded by long-ranged supercurrents, which 
decay on the lengthscale of the magnetic penetration depth $\lambda$. 
For the purpose of this work, these supercurrents are well approximated by the
vortex solution of London electrodynamics, since the exact behaviour
in the core region is of no relevance here.
In the following, the real gauge is employed, so that a phase gradient has already 
been absorbed into the magnetic vector potential $\mathbf A$, which itself is then proportional
to the superfluid velocity. The London solution of a single free Abrikosov vortex is given by
\begin{equation}
\label{EQAfree}
{\bf A}_v({\bf r},{\bf r}_v)=-\frac{\Phi_0}{2\pi \lambda} K_1 \left( \frac{ \left| {\bf r}-{\bf r}_v \right|}{\lambda}\right)\cdot 
\frac{(x-x_v) \hat{\bf y}-(y-y_v) \hat{\bf x}}{ \left| {\bf r}-{\bf r}_v \right|},
\end{equation}
where $\Phi_0=hc/2e$ is the flux quantum, $K_1$ a modified Bessel function of
first order, and  ${\bf r}=(x,y)$ the space coordinate. Accordingly, ${\bf r}_v=(x_v,y_v)$ denotes the
position of the vortex. The correct vortex solution for a geometrically restricted area 
has to fulfil Neumann boundary conditions, i.e., the superfluid velocity and thus the magnetic
vector potential  must be parallel to the boundaries. Generally, the free vortex according to
Eq. (\ref{EQAfree}), which represents a circular vector field around ${\bf r}_v$, does not accomplish this condition. 
For some simple geometries, however,
it is possible to construct the solution by placing virtual vortices and antivortices
outside the superconducting area, which is analogous to the technique of mirror charges
commonly known from classical electrostatics.
For the concrete case of the vortex lattice $\Lambda_0$ in a superconducting strip of width $L$ considered here, 
the correct solution can be generated by adding infinitely many virtual lattices $\Lambda_{n\neq 0}$, consisting
of (anti)vortices for $n$ even (odd).
A sketch of the situation is shown in Fig. $\ref{Fig02}$. Obviously,
this arrangement of vortices has the required symmetry of exchanging 
vortices and antivortices when mirrored at the boundaries 
$x=0$ or $x=L$, respectively, resulting in the correct implementation of the Neumann
boundary conditions. Thus, the magnetic vector potential
due to  the Abrikosov vortices can be written as
\begin{equation}
\label{EQAL}
{\bf A}_\Lambda ({\bf  r})=\sum_{n=-\infty}^{\infty}\sum_{{\bf r}_v \in \Lambda_n} (-1)^n {\bf A}_v({\bf r},{\bf r}_v).
\end{equation}
Note, that this contribution from the Abrikosov vortices corresponds to a magnetic field ${\bf B}_\Lambda=\nabla \times {\bf A}_\Lambda $ that
vanishes along the boundaries per construction. An external magnetic field ${\bf B}_0$ applied
along the $z$-axis has to be taken into account separately by the well-known Meissner solution of London theory: 
\begin{subequations}
\begin{eqnarray}
{\bf B}_M ({\bf r})&=&B_0 \frac{\cosh  (x/\lambda-L/2\lambda)}{\cosh(L/2\lambda)} \hat{\bf z}\\
\label{EQAM}
{\bf A}_M ({\bf r})&=& \lambda B_0 \frac{\sinh (x/\lambda-L/2\lambda )}{\cosh(L/2\lambda)} \hat{\bf y}.
\end{eqnarray}
\end{subequations}
Now that both the contributions of Abrikosov vortices and Meissner currents are found,
the total magnetic vector potential is just their superposition: ${\bf A}={\bf A}_M+{\bf A}_\Lambda$.

In the following, the zero-bias conductance peak split $\delta=\frac{e}{c}v_F |{\bf A}|$, 
which is expected at the boundary due to the Doppler shift of the superfluid velocity, will be calculated. 
Note, that the relevant quantity 
for the  tunneling measurements is  $\bar \delta$, which is spatially averaged  along the boundary. 
The Meissner contribution is constant along the boundary anyway, and the result at $x=0$  directly is
\begin{equation}
\label{EQAM}
{\bar{\bf A}}_M = - \lambda B_0 \tanh(L/2\lambda) \hat{\bf y}.
\end{equation}
The contribution of the vortex lattice $\Lambda_0$ and its mirror lattices $\Lambda_{n\neq 0}$ is periodic on the length $s$ (cf. Fig. \ref{Fig01}).
Therefore, their spatial average along the boundary is given as
\begin{equation}
{\bar{\bf A}}_\Lambda = \frac{1}{s} \int_0^{s} dy {\bf A}_\Lambda (x=0,y).
\end{equation}
It it possible to integrate out the $y$-dependence  analytically, which leads to
\begin{equation}
{\bar{\bf A}}_\Lambda = \frac{\Phi_0}{s}  \sum_{n=0} ^\infty  \sum_{x_v \in \Lambda_n} (-1)^n e^{-x_v /\lambda} \hat{\bf y}.
\end{equation}
The exponential term is  the contribution of all Abrikosov vortices with the same  $x_v$-coordinate, corresponding to a
whole dashed line in Fig. \ref{Fig01}. Indexing these positions $x_v$ according to $x_v=n L+m d+d_0$ with $m=0,1,...,m_c=(L-2d_0)/d$ leads to
 \begin{equation}
{\bar{\bf A}}_\Lambda = \frac{\Phi_0}{s} e^{-d_0 /\lambda} \sum_{n=0} ^\infty  \sum_{m=0} ^{m_c} (-1)^n e^{-n L /\lambda} e^{-md/\lambda} \hat{\bf y}.
\end{equation}
Analytic evaluation of both geometric series results in
 \begin{equation}
\label{EQALatt}
{\bar{\bf A}}_\Lambda = \frac{\Phi_0}{2s} \frac{\sinh(L/2\lambda+d/2\lambda -d_0 /\lambda)}{\cosh(L/2\lambda) \sinh(d/2\lambda)} \hat{\bf y}.
\end{equation}
At this stage, it is essential to employ an additional condition, which ensures the stability of the vortex lattice.
For a specific vortex of the lattice to remain stable, the current density generated by all other vortices, antivortices and
the Meissner screening current should vanish 
at the corresponding vortex position.
After a rather lengthy but straightforward analysis similar to the one above, the following condition can be derived
\begin{equation}
\label{EQStab}
\frac{\Phi_0}{2s}=\lambda B_0 \frac{\sinh (d/2\lambda)}{\cosh (d/2\lambda -d_0 /\lambda)},
\end{equation}
which guarantees the stability of all vortices simultaneously, i.e., the whole vortex lattice is stable. 
Plugging the stability condition Eq. (\ref{EQStab}) into Eq. (\ref{EQALatt}) yields
\begin{equation}
{\bar{\bf A}}_\Lambda = \lambda B_0 \left[\tanh(L/2\lambda)+ \tanh(d/2\lambda -d_0 /\lambda) \right]  \hat{\bf y}.
\end{equation}
Note, that the first term, which depends on the width $L$ and is linear in the applied magnetic field amplitude $B_0$, exactly cancels
the Meissner contribution ${\bar{\bf A}}_M$ of Eq. (\ref{EQAM}). Only the remaining second term of the vortex lattice contributes
to the total superfluid velocity, so that eventually
\begin{equation}
\bar{\delta}= \frac{e}{c} v_F \lambda B_0 \left| \tanh(d/2\lambda -d_0 /\lambda) \right|.
\end{equation}
It is important to realize, that this remaining term is not linear in $B_0$, since the parameters
$d$ and $d_0$ characterizing  the vortex lattice scale as a function of the magnetic field themselves.
However, because $d/\lambda,d_0/\lambda \ll1$ already for intermediate fields $B_0$,  Eq. (\ref{EQStab}) 
may be expanded in these small parameters, yielding $d= (q_\Lambda \Phi_0 /B_0)^\frac{1}{2}$. 
Accordingly, the final result for the zero-bias conductance peak split due to the Abrikosov vortices
is
\begin{equation}
\label{EQResult}
\bar \delta= \frac{h}{2 \sqrt{\Phi_0}} v_F f_\Lambda \sqrt{B_0}
\end{equation}
with the dimensionless number
$
f_\Lambda=\sqrt{q_\Lambda} \left|\frac{1}{2}- \frac{d_0}{d} \right|.
$

\begin{figure}[t]
\includegraphics[width=0.75 \columnwidth]{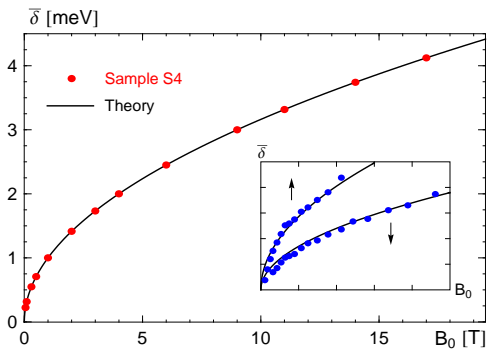}
\caption{\label{Fig03} (color online)
Zero-bias conductance peak split $\bar \delta$ as a function of the applied magnetic field $B_0$.
Main figure: Experimental data for sample S4 of Ref. \cite{Leibovitch} in decreasing fields (red points) and 
theoretical fit according to Eq. (\ref{EQResult}) with $f_\Lambda=0.0845$, which follows e.g. for a triangular 
Abrikosov vortex lattice with an offset of $d_0/d=0.5909$. Inset: Experimental data for
the 600 {\AA}  sample of Ref. \cite{Beck}, Fig. 3, in increasing and decreasing fields (blue points). Curves are
sample fits for a triangular vortex lattice at an offset of $d_0/d=0.6554$ (increasing field) and $d_0/d=0.5923$ (decreasing field). 
}
\end{figure}

The derived Eq. (\ref{EQResult}) is in complete agreement with the experimentally established universal relation Eq. (\ref{EQSqrtB}). 
Besides natural constants, the prefactor consists of the Fermi velocity $v_F$ as a material parameter
and the factor $f_\Lambda$, which is determined by the Abrikosov vortex lattice type and offset. 
Reasonable assumptions for the vortex lattice geometry allow a perfect fit to experimental data
by Beck et al. \cite{Beck,Leibovitch}, cf. Fig. \ref{Fig03}.  
Apart from the quantitative explanation of the universal magnetic field dependence,
the Abrikosov vortices might also provide some qualitative insight to the commonly 
observed hysteresis effects of the splitting. The number of Abrikosov vortices in a given
sample is higher, if the magnetic field $B_0$ is accessed downwards from 
higher magnitudes compared to the case, where the magnetic field has been
increased. For equal shape of the vortex lattice and equal intervortex distance, this directly yields
a larger offset parameter $d_0 /d$ at increasing magnetic fields and correspondingly a larger splitting, 
cf. the inset of Fig. \ref{Fig03}.
Finally, it should be pointed out that the resulting Eq. (\ref{EQResult}) does not 
depend on the width $L$ of the superconducting strip any more, as experimentally observed. 
Albeit speculative, even the one exceptional sample of different $\alpha$ and much smaller width 
might be explained. It is reasonable to assume, that a phase transition of the vortex lattice structure is 
induced geometrically by the further narrowing of the strip width, here. For example, an induced phase transition from
triangular to rotated triangular shape (cf. Fig. \ref{Fig01}) would lead to a different value of $q_\Lambda$ 
and result in an amplitude $\alpha$ different by a factor of $\frac{1}{\sqrt{3}}\approx 0.58$, which is close to the reported
value of $\frac{0.65}{1.1}\approx 0.59$.

This work has shown, that the long-ranged current contributions from Abrikosov vortices 
inside a $d$-wave superconductor are directly responsible for the universal magnetic field
dependence of the zero-bias conductance peak split observed experimentally.
Based on this derivation, key properties are fully explained and other 
observed features are accessible qualitatively.
Deviations of the average Abrikosov vortex density from that of a regular lattice, 
e.g. due to strong pinning, expected to result in deviations from Eq. (\ref{EQSqrtB}), 
are beyond the scope of this work. Nevertheless,
the zero-bias conductance peak split for arbitrary vortex distributions may
be derived similarly as presented here in a straightforward way.   

I acknowledge valuable discussions with N. Schopohl, T. Dahm, and M. Sigrist,
and I am grateful to R. Beck for providing me with the experimental results.
This work was financially supported by the Swiss Nationalfonds and the NCCR MaNEP.

\end{document}